\documentclass[useAMS,usenatbib]{mn2e}
\usepackage{graphicx}
\usepackage{rotating}
\usepackage{aas_macros}

\title[GMOS Spectroscopy of the S0 galaxy NGC 3115]{GMOS Spectroscopy of the S0 galaxy NGC 3115}
\author[Mark A. Norris, Ray M. Sharples and Harald Kuntschner]{Mark A. Norris$^{1}$,\thanks{E-mail:
m.a.norris@durham.ac.uk (MAN)} Ray M. Sharples$^{1}$, and Harald Kuntschner$^{2}$ 
\\
$^{1}$Department of Physics, Science Laboratories, South Road, Durham, UK\\
$^{2}$Space Telescope European Coordinating Facility, European Southern Observatory, Karl-Schwarzchild-Str. 2, 85748 Garching, Germany}
\begin{document}

\date{Accepted 2005 ***. Received 2005 ***; in original form ***}

\pagerange{\pageref{firstpage}--\pageref{lastpage}} \pubyear{2005}

\maketitle

\label{firstpage}

\begin{abstract}
We present Gemini GMOS longslit spectroscopy of the isolated S0 galaxy NGC 3115. We have determined kinematical data and Lick/IDS absorption line-strength indices for the major axis out to
around 9 kpc and for the minor axis out to around 5 kpc (around 2R$_e$). Using stellar population models which 
include the effects of variable [$\alpha$/Fe] ratios we derive metallicities, abundance ratios and
ages for the stellar population of NGC 3115. We find that [$\alpha$/Fe] remains
fairly constant with increasing radius at around [$\alpha$/Fe] = 0.17 for the major axis but increases rapidly for the minor axis to around [$\alpha$/Fe] = 0.3. We also find that to first order this behaviour can be explained by a simple spheroid + disc model, where the spheroid has [$\alpha$/Fe] = 0.3 and the disc shows close to solar abundance ratios. The disc
also appears considerably younger than the spheroid, having an age of around 6 Gyr
compared to 12 Gyr for the spheroid. We compare these results to those previously
presented for the globular cluster system of NGC 3115.
\end{abstract}

\begin{keywords}
galaxies: general - galaxies: abundances - galaxies: individual: NGC 3115 - galaxies: stellar content - globular clusters: general\end{keywords}

\section{Introduction}
Globular clusters (GC's) are among the simplest stellar systems. The stellar content
of individual clusters form primarily at one epoch and location and so are remarkably 
uniform in terms of metallicity, age and chemical abundances. Because 
GC's are believed to form preferentially 
during the major star forming and mass accumulation epochs, a careful 
examination of the GC population of a galaxy can shed light on these 
periods of a galaxies' development. 
This however relies on the assumption that GC's act as good tracers
of the properties of the overall stellar population formed at the same epoch.
For a more thorough discussion of why this is believed to be the case see 
\citet{Puzia05}. It is the aim of the present work to test this hypothesis in detail for one
well studied galaxy.

        NGC 3115 is one of the closest and most studied S0 galaxies. 
Its GC system has been extensively investigated, using both photometric \citep{Kavelaars,Kundu98,Puzia02}, and spectroscopic \citep{Kuntschner02,Puzia04} techniques. These studies find two GC sub populations of mean metallicities 
[Fe/H] $\simeq$ -0.37 and [Fe/H] $\simeq$ -1.36. \citet{Kuntschner02} find 
that both GC sub populations have ages which are consistent with a single
epoch of formation about 12 Gyr ago.
The observations of \citet{Kundu98} and \citet{Kavelaars} are consistent
with the metal-rich clusters being associated with a rapidly rotating thick disc
system and the slower rotating metal-poor clusters being associated with the halo
of NGC 3115.
The stellar population of the galaxy itself has also been examined, notably
 by \citet{Fisher96} who measured the line strengths and their gradients
out to a radius of 40" along the major axis, \citet{Trager98} also
examined the absorption line indices of the galaxy nucleus.
\citet{Elson97_n3115} found evidence for a bimodal distribution 
of metallicities with [Fe/H] $\simeq$-0.7 and [Fe/H] $\simeq$-1.3 in the resolved
stellar population of the halo located 8.5' east of the centre
and 5' from the major axis, although \citet{Kundu98} suggested that this
might be due to an instrumental effect in the metallicity calibration.
Taken together these observations lead to the possibility that the metal poor 
sub-population formed 12 - 13 Gyr ago, with the metal rich population forming 
several Gyr later after the ISM has been enriched by  a factor of $\simeq$4, perhaps 
triggered by AGN activity or a merger event.
        A further examination of the ages and metallicities of the general
stellar population and the GC population in this galaxy using updated Simple Stellar Population 
(SSP) models can test the plausibility of these scenarios.
 
        This paper is structured as follows. Section 2 describes the data set and its
reduction. Section 3 discusses the kinematical results of this investigation, Section 4
examines the line indices, and their use for calculation of luminosity weighted abundance ratios,
age and metallicity. Section 5 describes a simple two component fit to the [$\alpha$/Fe] ratios
of NGC 3115,  Section 6 includes a general discussion of the results of the 
previous sections and Section 7 presents our conclusions.

\begin{table}
 \centering
 \begin{minipage}{140mm}
  \caption{NGC 3115 Basic Parameters}
  \begin{tabular}{@{}llrrrrlrlr@{}}
  \hline
     Parameter    &  \multicolumn{2}{c} {Value} Source  \\
 \hline
Right Ascension (J2000)                 & 10$^{h}$05$^{m}$13.98$^{s}$ & ${a}$\\
Declination (J2000)                             & -07$^{\circ}$43{'}06.{"}9             & ${a}$\\ 
Morphological Type                              & S0$^{-}$                                      & ${a}$\\
Major Diameter                                  & 7.2 arcmin                                    & ${a}$\\
Minor Diameter                                  & 2.5 arcmin                                    & ${a}$\\
Heliocentric Radial Velocity            & 663   $\pm$4 \,kms$^{-1}$               & ${b}$\\
Asymptotic Radial Velocity              & 263 $\pm$5 \,kms$^{-1}$                 & ${b}$\\
Central Velocity Dispersion             & 314 $\pm$4 \,kms$^{-1}$         & ${b}$\\
Inclination of disc                             & 86$^{\circ}$                          & ${c}$\\
\hline
$^{a}$ NED. http://nedwww.ipac.caltech.edu/. \\
 $^{b}$ This study. \\
 $^{c}$ \citet{Capaccioli93}.
\end{tabular}
\end{minipage}
\end{table}

\section[]{Observations and Data Reduction}
The observations were carried out on the 18/19th of December 2001 with the GMOS 
instrument \citep{Hook04} on the Gemini North telescope (Program ID GN-2001B-Q-44). 
The B600 grism with 600 lines/mm
was used with a longslit 1 arcsec wide by 335 arcsecs long. The data were binned by 4 
in the spatial dimension and 2 in the spectral dimension producing a spectral resolution of 
$\sim$4.4\,\AA$ $ FWHM (110\,kms$^{-1}$) sampled at 0.9\,\AA/pixel. 
The seeing throughout the observations was generally $\le$ 0.8 arcsecs and the binned
pixel scale was 0.3 arcsecs/pixel. The wavelength range is $\sim$3800-6400\,\AA. 
Two sets of integrations were completed, one each for the major and minor axis 
of the galaxy, with a total integration time for each axis of 7200s. The centre of the major 
axis longslit was offset from the galaxy centre along the major axis by 120 arcsecs 
as can be seen in Fig. 1, to extend the radial coverage and reduce the galaxy light in the sky
background region. For the minor axis the sky region 
was defined at either end of the longslit. As can be seen from Fig. 1 the sky subtraction 
regions are located at large radii where the galaxy flux is only 5 - 6$\%$ of the sky level, 
hence over-subtraction of galaxy light should not be a major problem. This is discussed
in more detail in Sect. 2.1.
Two velocity standards HD97907, HD73665 and a photometric standard 
Feige 34 were also observed using the same experimental set up.

\begin{figure} 
   \centering
   \begin{turn}{-90}
   \includegraphics[scale=1.0]{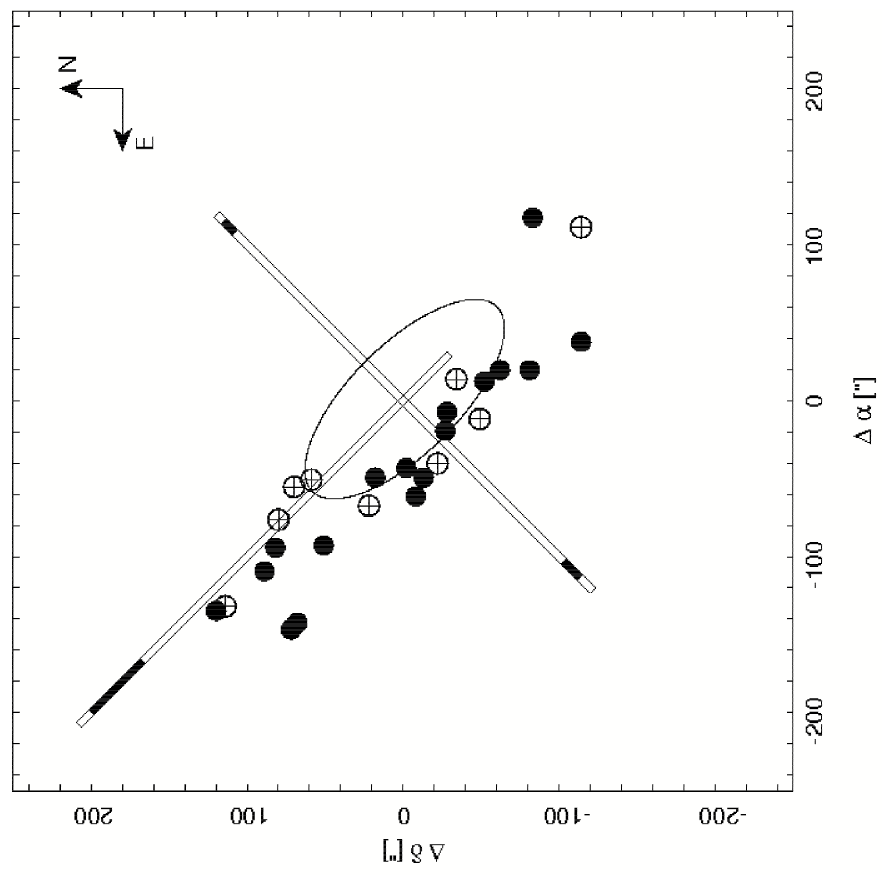}
   \end{turn} 
   \caption{Positions of the major and minor axis longslits. Slit widths exaggerated by a
   factor of 5. Dark regions define sky-subtraction regions. 
   Filled circles denote positions of GC's examined by 
   \citet{Kuntschner02} with line index measurements,
   unfilled circles have only kinematical data from \citet{Kuntschner02} available. 
   The ellipse indicates the position and orientation of the galaxy and encloses
   roughly half the integrated light \citep{Michard_n3115}.}
   \label{fig:1}
\end{figure}

The standard Gemini $\bf{IRAF}$ routines were used to carry out bias
subtraction, flat-fielding, and cosmic ray subtraction.  In addition to
these procedures it was noted that scattered light seriously
compromised the measurement of line indices at large galactocentric
radii. The scattered light had the effect of decreasing measured line
indices by applying a constant offset to the spectra; this effect leads to
the measurement of spurious age gradients at large radii. Fortunately
the GMOS set-up utilised in this investigation provided the opportunity
to remove this scattered light. This was possible because the 2-D
spectrum produced contained 3 unexposed regions created by the bottom
of the slit and two slit spacers located at 1/3 and 2/3 of the distance
along the spatial dimension of the image. After bias subtraction these
regions should contain no flux, however light scattered within the
spectrograph meant that this was not the case. To correct for this
effect it was possible to interpolate the scattered light level between
the 3 unexposed regions and to subtract this from the image. In total
scattered light accounted for a $\sim$ few $\%$ of the total incident
light. This however was sufficient to account for $\sim$ 50$\%$ of the
counts at the largest radii. After correction residual scattered light
should be of a negligible level.

        The data was then wavelength calibrated with the 
wavelength calibration being accurate to $\le$ 0.2\,\AA. The 2-D spectrum was then 
extracted into a series of 1-D spectra which were sky subtracted and binned 
in the spatial dimension until the target S/N (measured in a region near 
to the H$\it{\beta}$ line) of 20 or 60 (see Sect. 3) was reached. After binning to S/N = 60 
a total of 108 (major axis) and 44 (minor axis) spectra were produced, it is from these
spectra that all of the kinematics and line indices were measured, though in later figures
these have been rebinned in groups of 4 (radially) to allow clearer presentation. 
The binned 1-D spectra were then flux calibrated using the photmetric
standard and a reddening correction of E(B - V) = 0.146 \citep{Schlegel1998} 
applied using the standard $\bf{IRAF}$ routines.

 \begin{figure*}   
   \centering
   \includegraphics[scale=0.8]{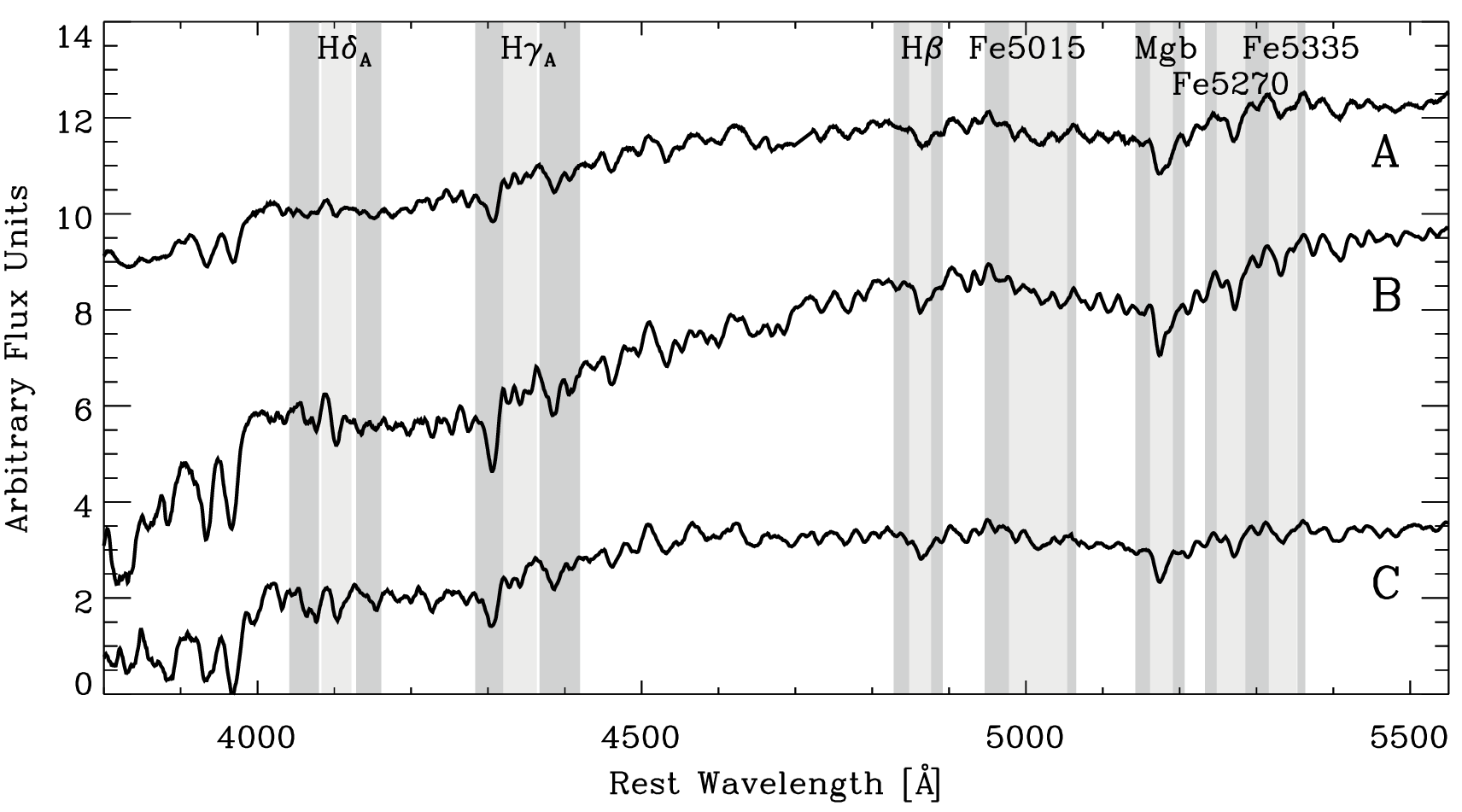}
   \caption{Typical rest-frame spectra corrected to the Lick/IDS resolution obtained in this study.
   Spectrum A is from the central region, Spectra B and C are the largest radii spectra from the major 
   ($\approx$ 130 arcsecs) and minor axes ($\approx$ 70 arcsecs) respectively. Over-plotted are the index bands (light grey) and red/blue 
   continua (dark grey) for the main Lick indices measured in this study.}
   \label{fig:2}
\end{figure*}

\subsection[]{Sky Subtraction}
Clearly an accurate sky subtraction is a necessity when attempting to
measure accurate
kinematics and
line indices at large galactocentric radii. If the sky spectrum is
strongly contaminated by diffuse stellar light it can introduce large
errors in measured indices and kinematic quantities such as $\sigma$,
h$_3$ and h$_4$. To investigate the magnitude of this effect several
methods were used, the simplest of which was to re-reduce our data
using the minor axis sky to sky subtract the major axis and vice versa.
The line indices were remeasured and compared to those measured
previously, for all indices considered here the changes in measured
index were $<$ 0.1\,\AA \, except for the Mg\,$b$\ line which showed
quite considerable variation. This change in the Mg\,$b$\ line can be
understood as being entirely due to changes in the strength of the
5200\,\AA \, sky emission feature between the major and minor axis
exposures.
        
        The second test carried out was to reduce the data assuming $\pm$ 20$\%$ errors
in the sky spectra, again any changes in the measured indices were $<$ 0.1\,\AA \, for
all cases except for the very last (and largest) radial bin. We are therefore confident 
that errors introduced by the sky subtraction should be entirely negligible for all but 
the very largest (and most binned) radial bins.

\subsection[]{Kinematics}       
        The first step in analysing the spectra was to measure the line 
of sight velocity distribution (LOSVD) for each of the binned spectra. The publicly 
available $\bf{IDL}$ implementation of the pPXF method \citep{pPXF} was employed 
to determine the recessional velocity, velocity dispersion, and the Gauss-Hermite 
moments $h_3$ and $h_4$ for the individual spectra. In this method the 
NGC 3115 spectra were fit in pixel space over the wavelength range 4900 - 5450\,\AA$ $ 
using both the stellar velocity templates observed during the run and additional stellar 
templates from the library of \citet{Vazdekis99}. Errors were estimated by remeasuring
 100 Monte-Carlo simulations of the spectra with added photon noise.
 Additionally, as a check of accuracy, 
the redshift and velocity dispersion were measured by Fourier cross-correlating 
the spectra against the velocity standards using the $\bf{fxcor}$ implementation
in $\bf{IRAF}$. As the two methods produce consistent results, the more
comprehensive pPXF implementation was used for the remaining analysis. 

\subsection[]{Line Indices}     
The system of line strength indices most commonly used for optical studies
is the LICK/IDS system \citep{Worthey97,Trager98}. We convolve our 
spectra with a wavelength dependent Gaussian kernel to reproduce the 
LICK/IDS resolution of $\sim$9 - 11\,\AA.

The line strength indices were then measured for each spectrum and corrected
for the smearing effects of the LOSVD following the procedure of \citet{Kuntschner04}. 
This procedure corrects the line index measurements for the effects of velocity
 dispersion and the non-gaussian $h_3$ and $h_4$ terms. \citet{Kuntschner04} finds that for
 changes of $\pm$0.1 in $h_4$ with constant $\sigma$ = 250 \,kms$^{-1}$ the LOSVD
 correction changes by $\pm$5$\%$ with corresponding errors in the age and metallicity
 estimations of 15 - 20$\%$. Clearly for a galaxy such as NGC 3115, with central
 velocity dispersion of around 300 \,kms$^{-1}$, the effect of these corrections on the
 spectra from the central few arcsecs could be non negligible. We therefore only measure
 indices for which \citet{Kuntschner04} has provided corrections for these effects (17 indices)
 though in practice we only make use of the H$\delta{_A}$, H$\gamma{_A}$, H$\beta$, Fe5015, Mg\,$b$\, 
 Fe5270 and Fe5335 indices in the following analysis. Errors in the line indices were calculated
 by 500 Monte - Carlo simulations of the input spectra.
  
 Because of the limited number of stars in the Lick library observed
 during the course of the observations, an accurate determination of
 the offsets \citep[see e.g.,][]{2000MNRAS.315..184K} between line
 indices measured here and the Lick system could not be determined
 empirically. However, we made use of the stellar library observed by
 \cite{Jones97} to determine the offsets for indices measured on flux
 calibrated spectra to the original Lick/IDS system assuming that the
 Jones library is well flux calibrated. The spectra were first
 broadened to the Lick/IDS resolution with a wavelength dependent
 Gaussian assuming a spectral resolution of the Jones library of
 1.8\,\AA~(FWHM). Then we compared the index measurements in common
 with the Lick stellar library (Worthey et al. 1994) for up to 128
 stars depending on the availability of the measurements. The offsets
 and associated errors (see Table~\ref{tab:lick_offsets}) were derived
 with a biweight estimator. The offsets are generally small
 ($<0.1$\,\AA) but individual indices can show larger offsets (e.g.,
 H$\gamma_{\rmn A}$, Fe5015).
 Here, we have only considered a single offset per index and ignored
 possible trends with line-strength. Several indices (e.g.,
 H$\delta_{\rmn A}$, CN$_2$, G4300) show weak evidence for such trends
 which are, however, difficult to quantify (see also Vazdekis 1999).
 Our determination of Lick offsets derived from Jones stars is in
 excellent agreement with an earlier investigation carried out by
 \citet[][Table 9]{Worthey97}. For the present study we apply the
 offsets listed in Table~\ref{tab:lick_offsets}.

\begin{table}
\caption{Offsets to the Lick/IDS system derived from Jones (1997) library}
\label{tab:lick_offsets}
\begin{tabular}{lcc} \hline
Index              & Offset           &   Number of stars  \\ \hline
H$\delta_{\rmn A}$ &$ -0.36 \pm0.05$\,\AA &108 \\
H$\delta_{\rmn F}$ &$ -0.16 \pm0.03$\,\AA &110 \\
CN$_1$             &$  0.003\pm0.002$\,mag&117 \\
CN$_2$             &$  0.006\pm0.002$\,mag&115 \\
Ca4227             &$  0.00 \pm0.02$\,\AA &126 \\
G4300              &$ -0.28 \pm0.04$\,\AA &128 \\
H$\gamma_{\rmn A}$ &$  0.38 \pm0.04$\,\AA &126 \\
H$\gamma_{\rmn F}$ &$  0.09 \pm0.02$\,\AA &126 \\
Fe4383             &$  0.28 \pm0.05$\,\AA &126 \\
H$\beta$           &$ -0.12 \pm0.02$\,\AA &128 \\
Fe5015             &$  0.23 \pm0.04$\,\AA &126 \\
Mg\,$b$            &$ -0.08 \pm0.02$\,\AA &128 \\
Fe5270             &$ -0.07 \pm0.02$\,\AA &128 \\
Fe5335             &$ -0.04 \pm0.03$\,\AA &128 \\
Fe5406             &$ -0.06 \pm0.02$\,\AA &126 \\
\hline
\end{tabular}


Notes: Column (1) gives the index name, while column (2) gives the mean
offset (Lick - Jones) to the Lick/IDS system evaluated from the Jones
(1997) stars in common with Lick. Column (3) shows the number of stars
used in the comparison.


\end{table}

 \begin{figure*}   
   \centering
   \includegraphics[scale=0.9]{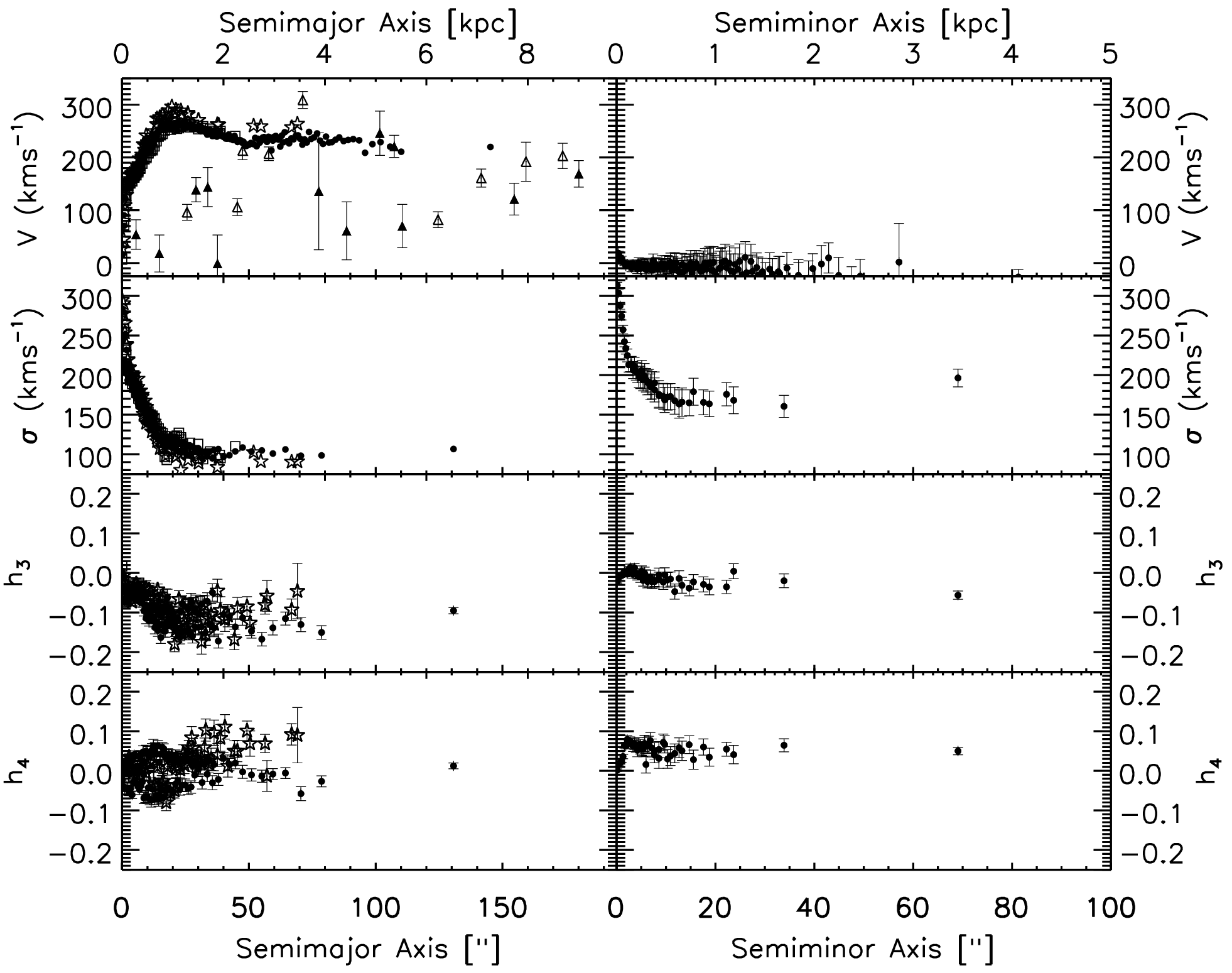}
   \caption{Comparison of measured LOSVD parameters with literature values.
   Filled circles are from this study, asterisks from \citet{Bender94}, and
   squares from \citet{Fisher97}. Triangles show GC data from \citet{Kuntschner02}
   de-projected to the major axis; filled triangles 
   are the blue sub-population and unfilled
   the red sub-population. Several GC points are omitted because they exhibit counter
   rotation or have implied rotation velocities of greater than 400\,kms$^{-1}$.
   Boot-strapped 1$\sigma$ error bars are over-plotted for values determined in this
   study, except for major axis velocity and velocity dispersion data where they are omitted 
   in the interests of clarity. Typical errors in V and $\sigma$ are $\le$ 12 \,kms$^{-1}$. The 
   major axis data has been mirrored about the minor axis and about the central recessional velocity,
   the minor axis has similarly been mirrored about the major axis.}
   \label{fig:3}
\end{figure*}

\section[]{Galaxy Kinematics}
Fig 3 shows the result of the kinematical measurements, the velocity measurements are 
based on spectra binned to have S/N = 20, the other parameters were derived from spectra 
binned to have S/N = 60.  As a check that changes in S/N do not affect the
measured quantities, the data was re-binned to S/N of  30, 40 and 50; the kinematics
were then remeasured with no significant trends in measured quantities being observed.
        
        We shall discuss each of the parameters measured here in turn but in general 
the measured kinematical data is in very good agreement with published 
data from several authors including; \citet{Kormendy92}, \citet{Capaccioli93}, 
\citet{Bender94} and \citet{Fisher97} in almost all respects except for the 
inferred $h_4$ value and the rotational velocity at large radii.

The inner rotation curve of the major axis measured here is in 
good agreement between all the data sets we have 
examined including \citet{Illingworth82}, \citet{Kormendy92}, \citet{Capaccioli93}, 
\citet{Bender94}, \citet{Fisher97} and  \citet{Emsellem99}. At larger radii however
other authors including \citet{Capaccioli93} have measured an essentially flat
rotation curve with rotational velocity around 260$kms^{-1}$. We however observe some
evidence for some drop off beyond 70 arcsecs. As observed by several
authors, we find no evidence for minor axis rotation. Statistics are insufficient at present
to determine if either of the GC sub-populations can be better associated with
structures such as discs or spheroids within NGC 3115. It is however clear that the
bulk of the clusters examined rotate in a manner consistent with that of the bulk of the
stellar content even at larger radii. In fact of the 26 clusters with kinematics examined
here 21 rotate in a prograde manner.

\begin{table}
\caption{Major Axis Kinematics}
\begin{tabular}{|r|r|r|r|r|}
\hline
  \multicolumn{1}{|c|}{Radius ["]} &
  \multicolumn{1}{c|}{Velocity [km/s]} &
  \multicolumn{1}{c|}{$\sigma$ [km/s]} &
  \multicolumn{1}{c|}{H$_3$} &
  \multicolumn{1}{c|}{H$_4$} \\
\hline
  -130.71 & 229.33 & 106.66 & -0.098 & 0.013\\
  $\pm$ & 4.15 & 5.49 & 0.008 & 0.007\\
  -78.66 & 245.07 & 98.98 & -0.158 & -0.028\\
  $\pm$ & 7.9 & 8.98 & 0.017 & 0.014\\
  -70.52 & 244.13 & 98.48 & -0.140 & -0.062\\
  $\pm$ & 8.04 & 10.53 & 0.018 & 0.017\\
  -64.41 & 242.45 & 106.53 & -0.122 & -0.008\\
  $\pm$ & 8.29 & 9.1 & 0.017 & 0.013\\
  -59.47 & 242.97 & 101.13 & -0.147 & -0.012\\
  $\pm$ & 8.68 & 8.88 & 0.019 & 0.013\\
  -55.11 & 242.83 & 104.85 & -0.175 & -0.018\\
  $\pm$ & 7.97 & 9.24 & 0.017 & 0.014\\
\hline\end{tabular}
Notes: Table continued in electronic format. Tables 4 (minor axis kinematics), 
5 and 6 (major and minor axis line strengths, ages, [$\alpha$/Fe] and [Z/H])
are provided in electronic form at http://star-www.dur.ac.uk/~dph3man/data.html.
All data provided is binned to S/N = 60.
\end{table}

The velocity dispersion of NGC 3115 is also in good agreement with 
other measured results from the authors cited 
previously. However the data we present here extends to significantly 
larger radii than previous studies. One obvious 
feature of the data presented in Fig. 3 is that the minor axis 
velocity dispersion is considerably higher than that of the major axis at
equivalent radii. 
The fact that the minor 
axis data displays a higher velocity dispersion than the major axis is 
not entirely unexpected, as data presented by \citet{Kormendy92} hints 
at this being the case. The 2-D spectroscopy presented in \citet{Emsellem99} 
also displays evidence for lines of constant velocity dispersion being 
elongated in the minor axis direction (at least within the inner 5 arcsecs).
A difference in measured velocity dispersion for the major and minor axes
can be explained if the galaxy consists of a fast rotating, kinematically cool
disc component and a slower rotating, kinematically hot spheroidal component. 
This possibility is re-examined later in the light of line index measurements.

        The values of $h_3$ determined here are entirely consistent with values 
determined previously.

        The $h_4$ values measured here are generally consistent with those measured 
previously except at larger radii, where we observe a value $h_4$ of 0.0, 
but \citet{Bender94} find a value of around 0.1 this discrepancy could be due to several 
factors including differing experimental set-up and method of measuring h3 and h4.

\section[]{Line Indices}

\begin{figure*} 
   \centering
   \includegraphics[scale=0.9]{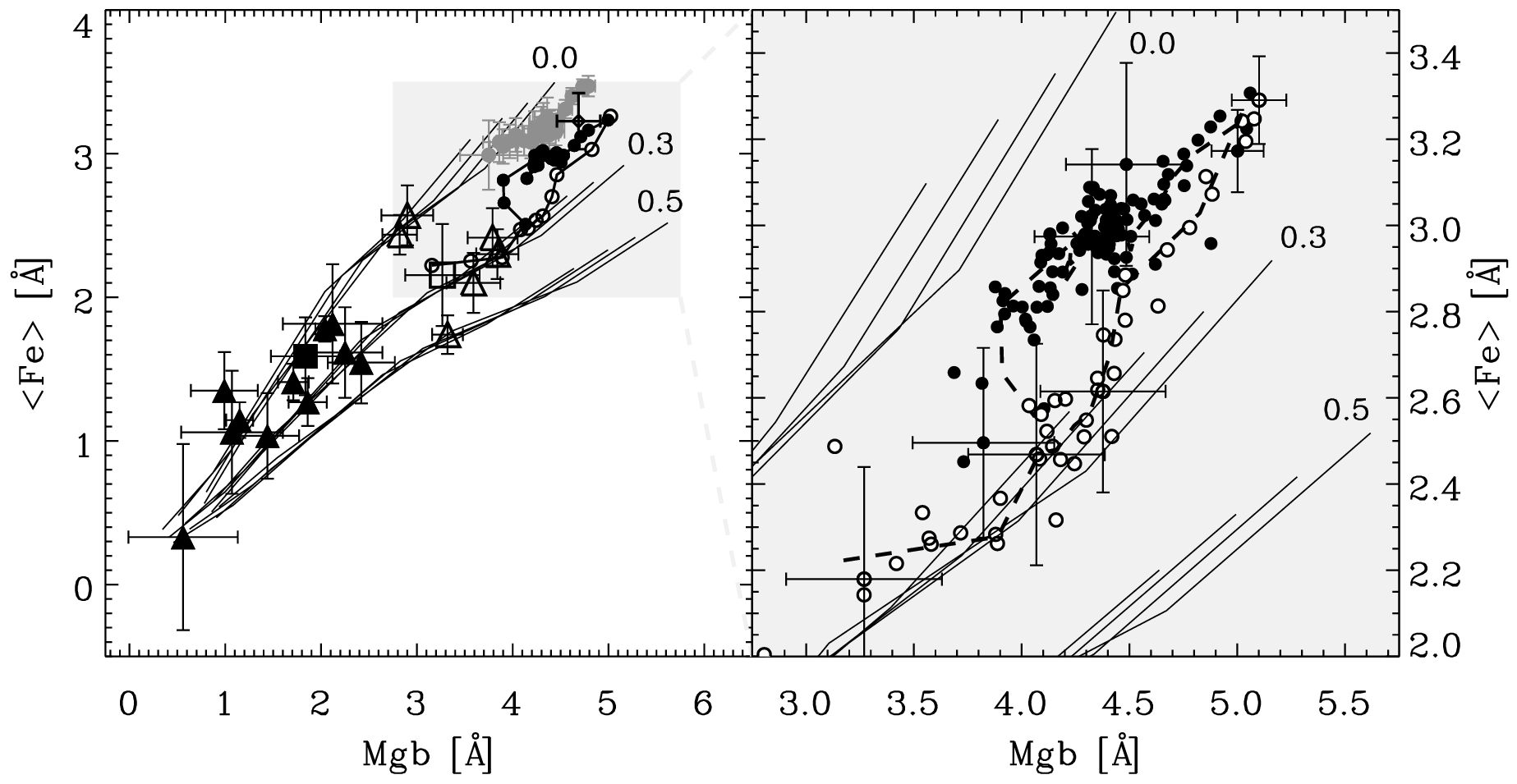}
   \caption{Comparison of the [$\alpha$/Fe] ratios of the galaxy data and GC population
   through the use of an Mg$\it{b}$ vs. $<$Fe$>$ diagram. In the left panel filled
   and open black circles are binned major and minor axis galaxy data from this 
   paper, filled grey circles show results from 
   \citet{Fisher96}, the open diamond is the central galaxy result from \citet{Trager98}. 
    Filled black triangles are the blue globular cluster sample and 
    unfilled triangles are the red globular cluster sample, both from
   \citet{Kuntschner02}. Filled black square is the error weighted mean of the blue
   cluster subpopulation and the black square the error weighted mean of the red
   cluster subpopulation.
   Overplotted are models by \citet{Thomas03,Thomas04} 
   with abundance ratios [$\alpha$/Fe] = 0.0, 0.3, 0.5, the models have ages 3-12 Gyr
   and metallicity [Z/H] = -2.25, -1.35, -0.33, 0.0 and +0.35.
   Right panel shows the unbinned galaxy data presented here, dashed lines
   denote the positions of the binned data. Representative error bars are plotted
   for spectra located at small, intermediate and large radii, errors are a combination of 
   bootstrapped 1$\sigma$ errors and those introduced by a $\pm$5$\%$ 
   error in the sky subtraction.}
   \label{fig:4}
\end{figure*}

\subsection[]{Abundance Ratios}
In Fig. 4 we show Mg$b$ vs. $<$Fe$>$ where $<$Fe$>$ = ((Fe5270 + Fe5335)$ $ / 2)  \citep{Gonzalez93}, for the major and minor axes of NGC 3115. The GC data points are
separated into red and blue sub-populations by (V - I) colour.  Model predictions
from \citet{Thomas04} are overplotted for [$\alpha$/Fe] of 0.0, 0.3  and 0.5
with ages = 3 (left most line in each group), 5, 8 and 12 (right most line in each group)
 Gyr and have [Z/H]  which range from -2.25 (bottom left)
 to +0.35 (top right). The effects of age and metallicity
are essentially degenerate in this diagram, with sensitivity to abundance
ratios maximised. As described in \citet{Kuntschner02}, abundance ratios are
most accurately determined for larger ages/metallicities.

The measurements for the centre of NGC 3115 from the major axis data
are in reasonable agreement with those from \citet{Fisher96} and also
from \citet{Trager98}. The implied $\alpha$ - element over-abundance
of [$\alpha$/Fe] $\approx$ 0.17 was determined by chi$^2$ minimization
of the data and the models of Mg$b$ vs. $<$Fe$>$ from \citet{Thomas04}.
This chi$^2$ procedure, which is similar to the one introduced by \citet{2004MNRAS.355.1327P}
fits the set of indices used here (H$\beta$, Fe5015, Mg\,$b$\, Fe5270
and Fe5335) to the models of \citet{Thomas04}, allowing variable ages,
[$\alpha$/Fe] and [Z/H]. A possible break in the value of
[$\alpha$/Fe] is visible for Mg$b$ $<$ 4 with an increase in
[$\alpha$/Fe] to around 0.25. In comparison, the minor axis data starts
off at values consistent with the central portion of the major axis,
and then begins to trend off towards [$\alpha$/Fe] = 0.3 much more
rapidly.

A break in the major axis data and differences between the two axes 
can be interpreted as evidence for the existence of
at least two distinct populations, with typical values of [$\alpha$/Fe] $\sim$ 0.17 
and 0.3. Since a move towards lower Mg$b$ and $<$Fe$>$ corresponds
to a move to increasing radius it raises the possibility of observing radial trends
in the strengths of other line indices, which will be examined in more detail in the 
next section.

Our data extends to sufficiently large radii that the mean metallicity of the stellar 
population is similar to that of the most $\it metal $ $ rich$ GCs. Fig. 4
shows that on both the major and minor axis the mean abundance ratios at the
largest radii are also consistent with those of the cluster population although
the spread in abundance ratios appears to be smaller. No significant population
of GCs are found with properties similar to the stellar population at intermediate 
radii on the major axis.

\subsection[]{Radial Profiles of Indices} 
Fig. 5. displays the radial profiles of the measured Lick indices, [$\alpha$/Fe]
and [Z/H]. The values of r$_e$ used here are those listed in \citet{Capaccioli93} 
for the spheroidal component of the galaxy.

        The differences in [$\alpha$/Fe] between major
and minor axes is more clearly demonstrated here, as it is evident that
at larger radii the major axis data again becomes consistent with 
that of the minor axis. This type of behaviour could be understood in terms
of changes in the relative contributions of disc and spheroidal components,
with both major and minor axes being dominated at small radii by a nuclear 
component. At intermediate radii the major axis would be affected by the influence
of the disc component, whereas the minor axis would simply be tracing the
stellar content of the spheroidal component. The change in the behaviour 
of the major axis at large radii could then be understood as evidence for
truncation of the disc component contribution at around 80 arcsecs on the
major axis.

        The other metallicity tracing indices (Mg\,$b$\, Fe5270 and Fe5335) display 
similar trends, with slight evidence for breaks 
in the major axis data at around the same radius as the one seen in [$\alpha$/Fe].
This has previously been observed by \citet{Fisher96} in their edge-on sample
of S0 galaxies (unfortunately they did not examine the minor axis of NGC 3115), 
who found that the Mg$_2$ index is stronger with a
lower gradient at larger radii on the major axis than on the minor axis. As can be
seen from Fig. 4 the values for Mg\,$b$\, Fe5270 and Fe5335 determined here
for the major axis are in good agreement with those determined by 
\citet{Fisher96}. Small differences between the two datasets can be attributed to variations
in the experimental set-up, the different methods employed to correct for
the broadening effect of the LOSVD, and the uncertainty on our correction to
the Lick system.

\begin{figure} 
   \centering
   \includegraphics[scale=0.9]{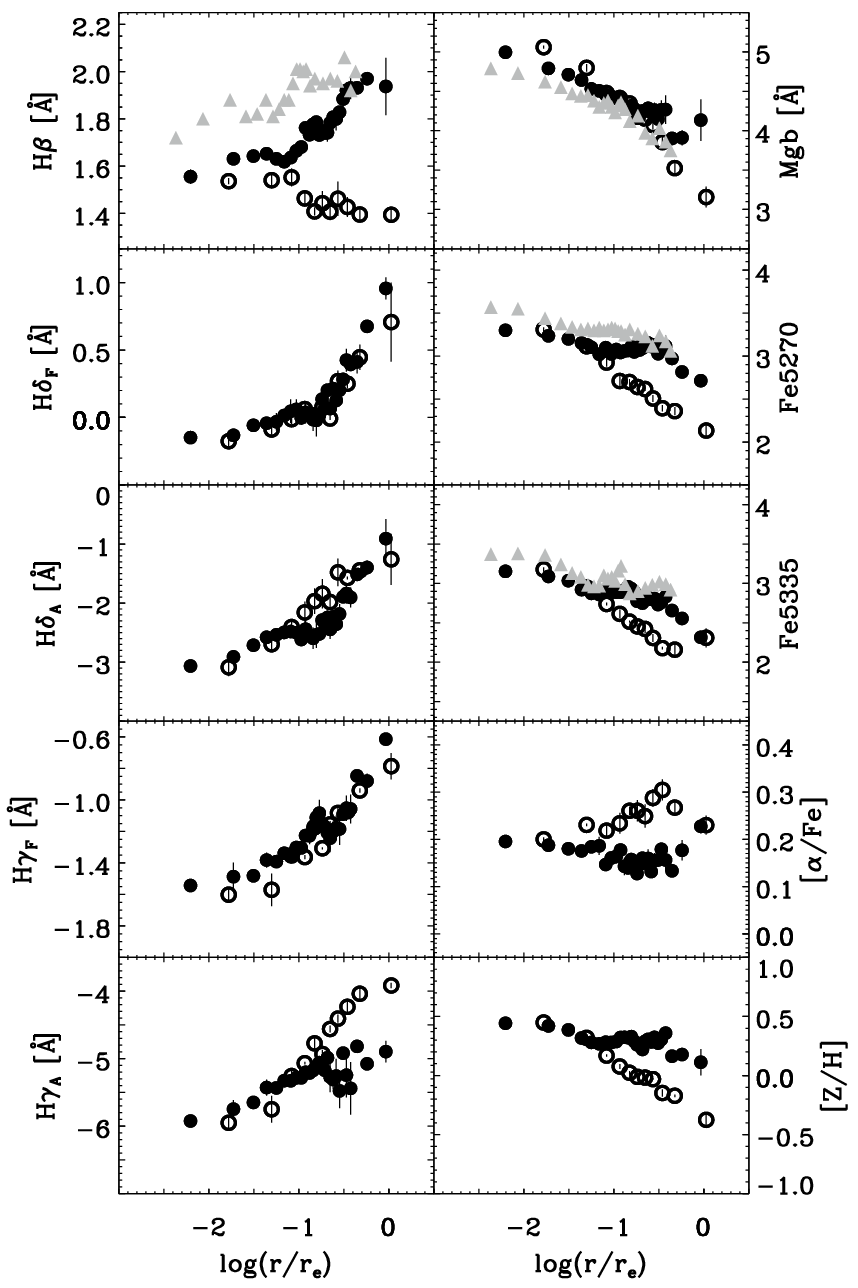} 
   \caption{Radial profiles of Lick Indices, [Z/H] and [$\alpha$/Fe]. [Z/H] and [$\alpha$/Fe] determined
   by chi$^2$ minimisation as described in Sect. 4.1.
   Filled black circles
   show major axis data, open circles show minor axis data and light grey triangles show data
   from \citet{Fisher96}. Error bars are 1-$\sigma$ errors in each bin. Note that r$_e$ = 93 and 35 arcsecs
   for the major and minor axes respectively \citep{Capaccioli93}.}
   \label{fig:5}
\end{figure}

\subsection{Age Determinations}
Fig. 6. examines the behaviour of age-sensitive indices along the major and minor
axes of NGC 3115.  Model grids are interpolated between the [$\alpha$/Fe] = 0.0 and 0.3 models
from \citet{Thomas03,Thomas04} with [$\alpha$/Fe] determined as described 
previously. Note that whilst the [$\alpha$/Fe]
ratios used to produce the major axis grids are appropriate for the longslit data described here,
they are generally not appropriate for the GCs which tend to have mean
[$\alpha$/Fe] $\sim$ 0.3. However the minor axis grids provide a good approximation
 to the mean [$\alpha$/Fe] of the GC's and hence the best age determinations for the clusters. 

        The H$\beta$, H${\gamma}_A$ and H${\delta}_A$ indices are
 plotted against [MgFe]$'$ where [MgFe]$'$ = $\sqrt{Mgb \times(0.72 \times Fe5270 +
 0.28 \times Fe5335)}$ \citep{Thomas03}. This index was found by \citet{Thomas03}
 to be independent of [$\alpha$/Fe] and a good tracer of total metallicity.

        There appears to be a difference in age between the two axes with the major
axis having a mean age of around 5 - 8 Gyr and the minor axis an age of 
around 12 Gyr. This would suggest a small amount of star formation may
have continued in the disc component for several Gyr after the formation of the
spheroid.

        The age determinations from the H$\beta$ index appear to change at lower [Z/H] 
but in the opposite sense for each axis, with the major axis appearing to become 
younger and the minor axis older.  As H$\beta$ is relatively unaffected by changes 
in [$\alpha$/Fe] this cannot be explained as being due to changes in [$\alpha$/Fe] 
along either axis, but could be explained for the major axis by more recent star 
formation in the outer parts of the disc (in spiral arms perhaps). 
Another possibility would be that some undetected H$\beta$ emission in the inner
regions of the disc weakens the observed H$\beta$ index in the inner regions. This
final possibility however seems unlikely since we find negligible signs of [OIII] emission.
For the minor axis data it would seem plausible that the populations at larger radii could
be older  and would in fact represent the older generation
of stars also being traced by the GC populations. This possibility would seem to gain
credence from the fact that at larger radii the minor axis data displays [$\alpha$/Fe], ages and
[Z/H] values that are entirely consistent with those determined from the red GC sub-population.

        The remaining age estimators H${\gamma}_A$ and H${\delta}_A$ also show an age
offset between the minor and major axis in the same sense as H$\beta$, but there is little or
no evidence for an age gradient.

        Fig.7. shows the radial profiles of ages determined by the chi$^2$ 
minimisation described previously. As should be expected the trends 
previously described are obvious with the major axis appearing to
have an age of 5 - 8 Gyr and the minor axis having an age of between 12 - 14 Gyr.

        A further comment is that despite the use of different age sensitive indices 
our age determinations for the GC's are entirely consistent with those 
of \citet{Kuntschner02} with both GC populations having a mean age of around 12 Gyr.
In fact the agreement is now improved as the mean ages of the sub-groups determined
by \citet{Kuntschner02} varied from  $\sim$ 6 Gyr to 12 Gyr, depending on the Balmer
line being examined. This spread can now be understood as being due to the [$\alpha$/Fe] sensitivity
of the H${\gamma}_F$ and H${\delta}_F$ lines used in the Kuntschner analysis, which can now
be corrected using the newer SSP models provided by \citet{Thomas04}.

\begin{figure*} 
   \centering
   \includegraphics[scale=0.7]{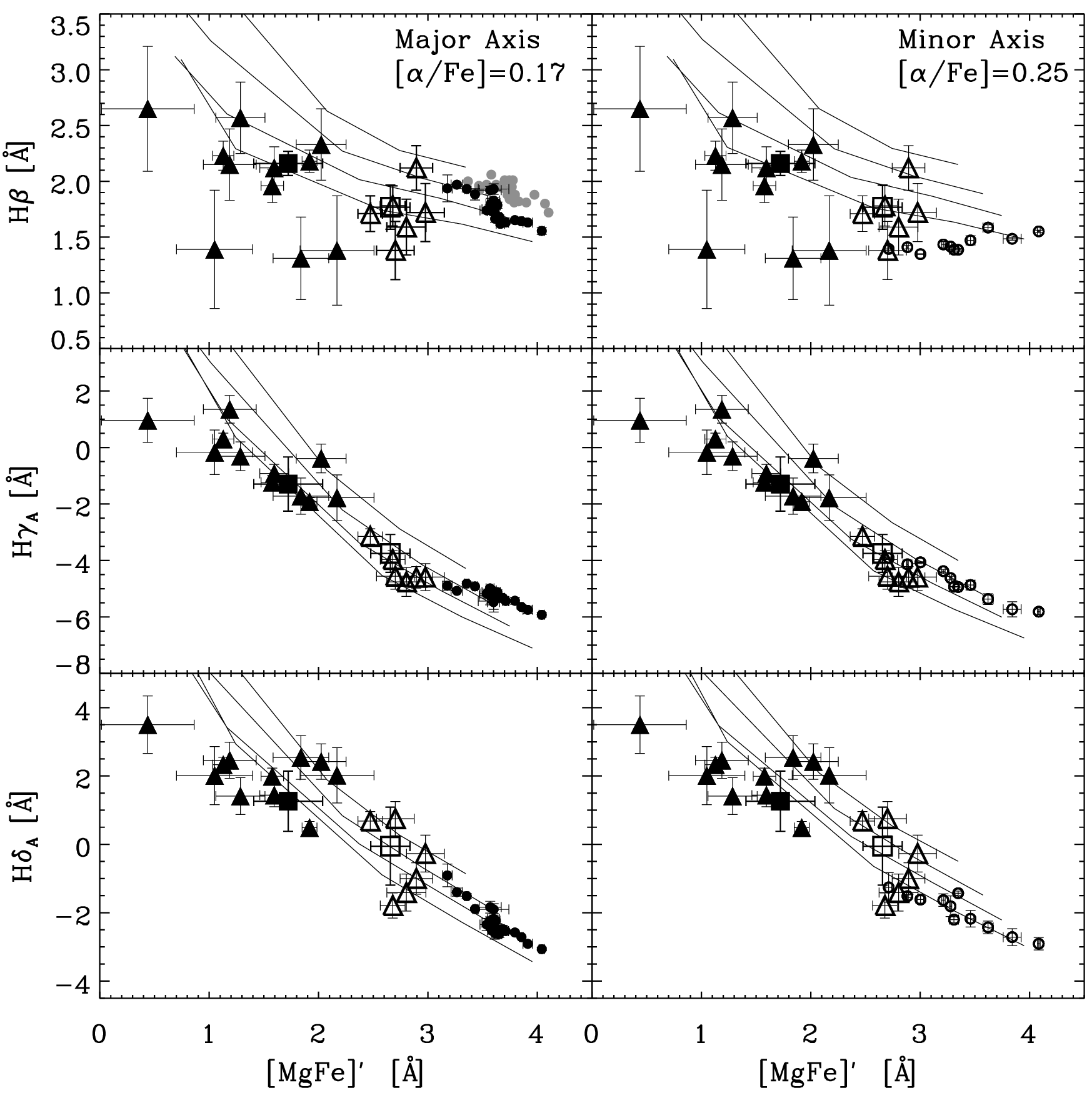}
   \caption{Age-metallicity diagnostic plots for major and minor axis data. Symbols are
   as in Figure 4. Overplotted are models by \citet{Thomas03,Thomas04} 
   with abundance ratios [$\alpha$/Fe] indicated in the top right corner of each
   column, a value of [$\alpha$/Fe] = 0.25 was chosen for the minor axis as an 
   average of the values determined at small and large radii. These [$\alpha$/Fe]  
   ratios are interpolated from the 0.0 and 0.3 values listed in \citet{Thomas04}, the 
   models have metallicity [Z/H] = -2.25 to +0.35, lines indicate ages
   from top to bottom of 3, 5, 8 and 12 Gyr respectively. Error bars for galaxy data are the
   1-$\sigma$ errors on each of the radial bins. }
   \label{fig:6}
\end{figure*}

\section[]{A Simple Two Component Model}
To test the hypothesis that radial trends in [$\alpha$/Fe] could be explained
by intrinsic differences in the disc and spheroid components a simple 
model was constructed.
        
        The relative contributions of the disc and spheroid components were
determined from archived GMOS images of NGC 3115 obtained 
in 2004 (Program ID GS-2004A-Q-9, PI R.M.Sharples) in the $\it{g}$,  $\it{i}$ and $\it{r}$ bands. 
The program GALFIT \citep{Peng02_Galfit} was used to carry out a simple bulge
 to disc decomposition. Because we only wish to know the relative contribution 
 of the two components and we are only interested in a simple first order model, 
 a full isophotal decomposition was unnecessary and a simple de Vaucouleurs 
 model of the spheroid with r$_e$ = 93 arcsecs \citep{Capaccioli93} was subtracted 
 from the original image to isolate the disc light contribution.
As a test of this approach, we also allowed GALFIT to attempt to fit the images
with a Sersic function with variable r$_e$ and $\it{n}$, both with and without masking
of the disc region and with varying starting parameters. The results in terms of
r$_e$ and $\it{n}$ varied considerably but the distribution of flux 
in the residual image remained fairly constant, with the disc tending to
provide a peak of around 30 - 40$\%$ of the flux on the major axis.

        The results of this bulge subtraction can be seen in Fig. 8. As previously
noted by  \citet{Capaccioli88} the disc
shows considerable flaring in the outer regions, which these authors attributed 
to the disc ceasing to be self gravitating in this region. There is also evidence for 
structure within the disc (spiral arms?), which could be taken as evidence for some residual 
star formation events. Note that the inner region is not well
fit by this model. A more realistic model would require several components, but is 
beyond what is required for the present analysis.

        The model can be used
to predict [$\alpha$/Fe] at any point, if it is assumed that the residual
light traces an enriched disc and nuclear component with [$\alpha$/Fe] $\sim$ 0.0
and that the fitted spheroid traces a lower Fe - enriched population of 
[$\alpha$/Fe] $\sim$ 0.3. By weighting the [$\alpha$/Fe] value by the relative fractions of the two components
it is possible to estimate the observed [$\alpha$/Fe] at any point. Fig. 9 shows the result of this 
procedure for both axes. This simple model reproduces the general trends observed
with [$\alpha$/Fe] quickly rising on the minor axis and a much more Fe-enriched major axis
which trends back towards the asymptotic values of the minor axis at large radii. 
The differences between the model and observed values can be explained 
by a number of factors including incorrect values for the intrinsic [$\alpha$/Fe] of the 
two components and the effect of other unaccounted for components.
 Other factors such as intrinsic gradients in [$\alpha$/Fe] could also play a part. However to first
order we believe that differences in [$\alpha$/Fe]  between the two axes of NGC 3115
can be explained as being due to the existence of at least two distinct stellar
populations within the galaxy, with different spatial distributions and enrichment histories.

\begin{figure} 
   \centering
   \includegraphics[scale=0.7]{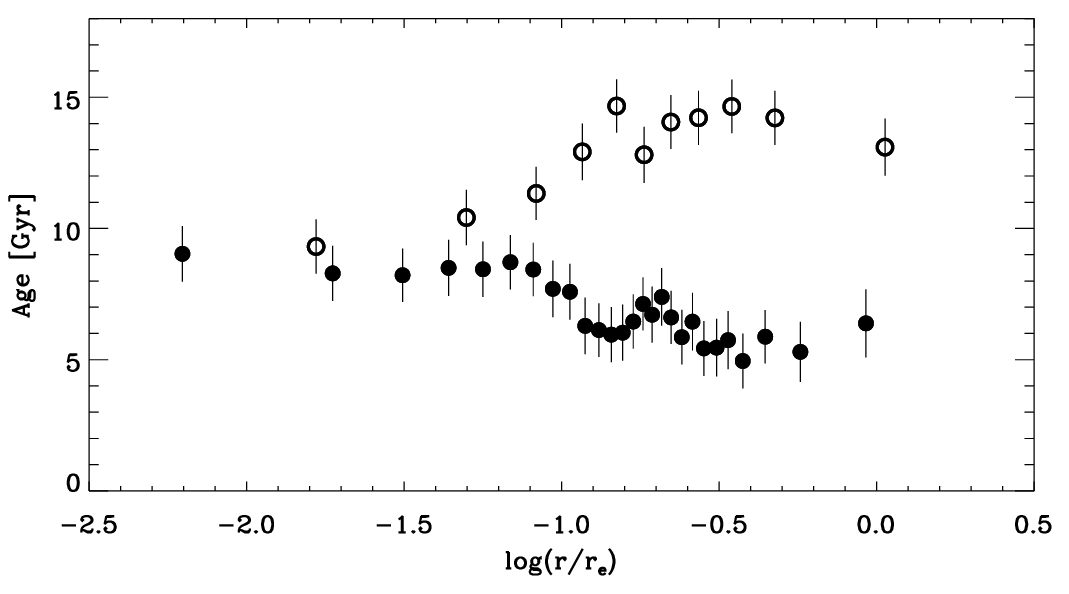} 
   \caption{Radial profiles of measured age for major and minor axes. Ages determined by
   chi$^2$ minimisation as described in Sect. 4.1.
   Symbols as previously defined.}
   \label{fig:7}
\end{figure}

\begin{figure} 
   \centering
   \includegraphics[scale=0.7]{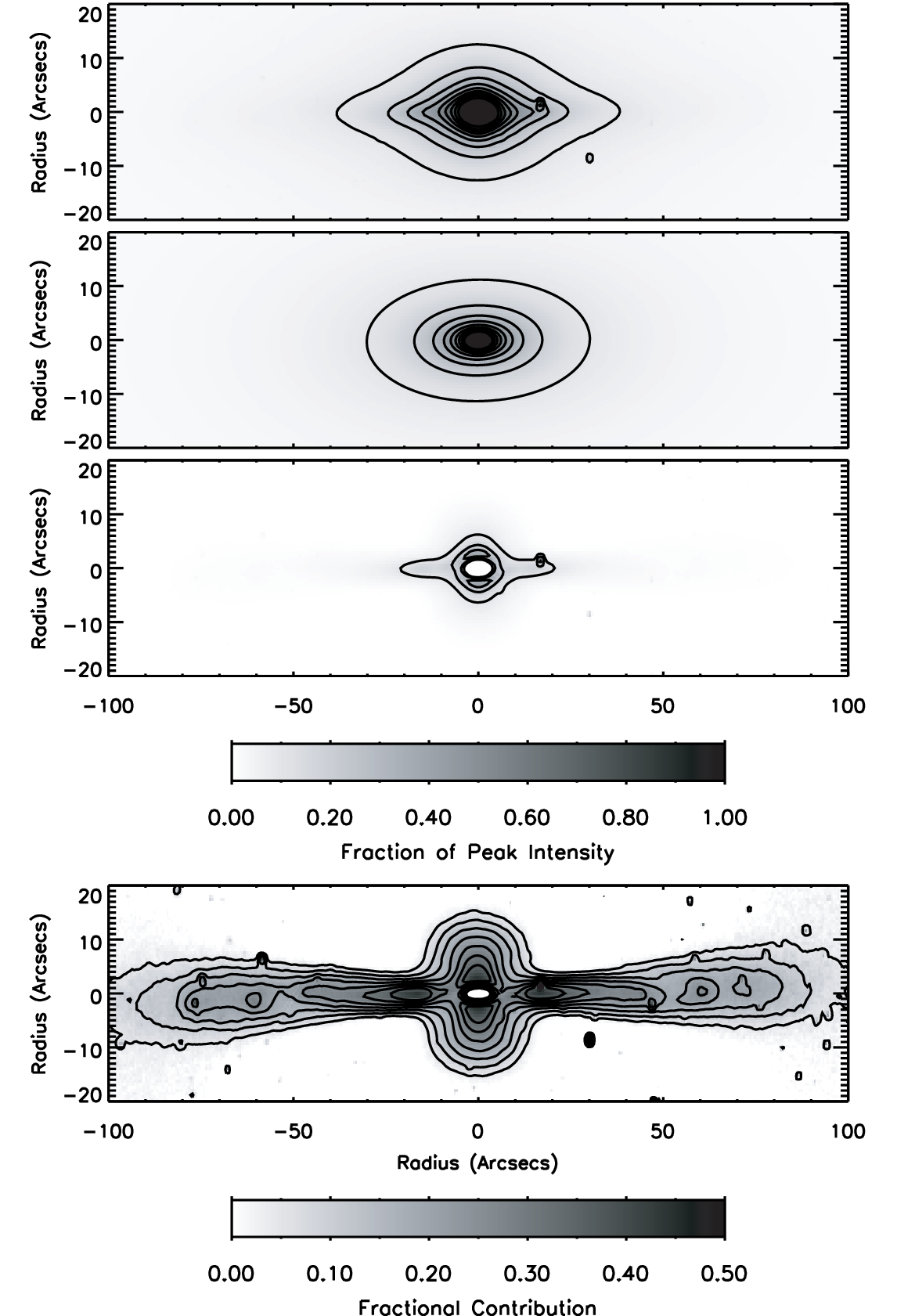} 
   \caption{Top image shows original GMOS $i$ band image. The 2nd image
   shows the simple de Vaucouleurs model for the spheroidal component.
   The 3rd image shows the residual image after the 2nd is subtracted from the
   1st. These 3 images are all scaled to the peak intensity of the original image
   for clarity. Contours show the 0.1, 0.2, 0.3, 0.4, 0.5, 0.6, 0.7, 0.8, 0.9 fractions. 
   The lower image shows the fraction of the total emission provided by
   the residual components. Considerable flaring of the disc is visible, as is
   substructure consistent with spiral arm structures. Contours show the 0.1, 0.2, 0.3, 
   0.4 and 0.5 fractions.}
   \label{fig:8}
\end{figure}

\begin{figure} 
   \centering
   \includegraphics[scale=0.5]{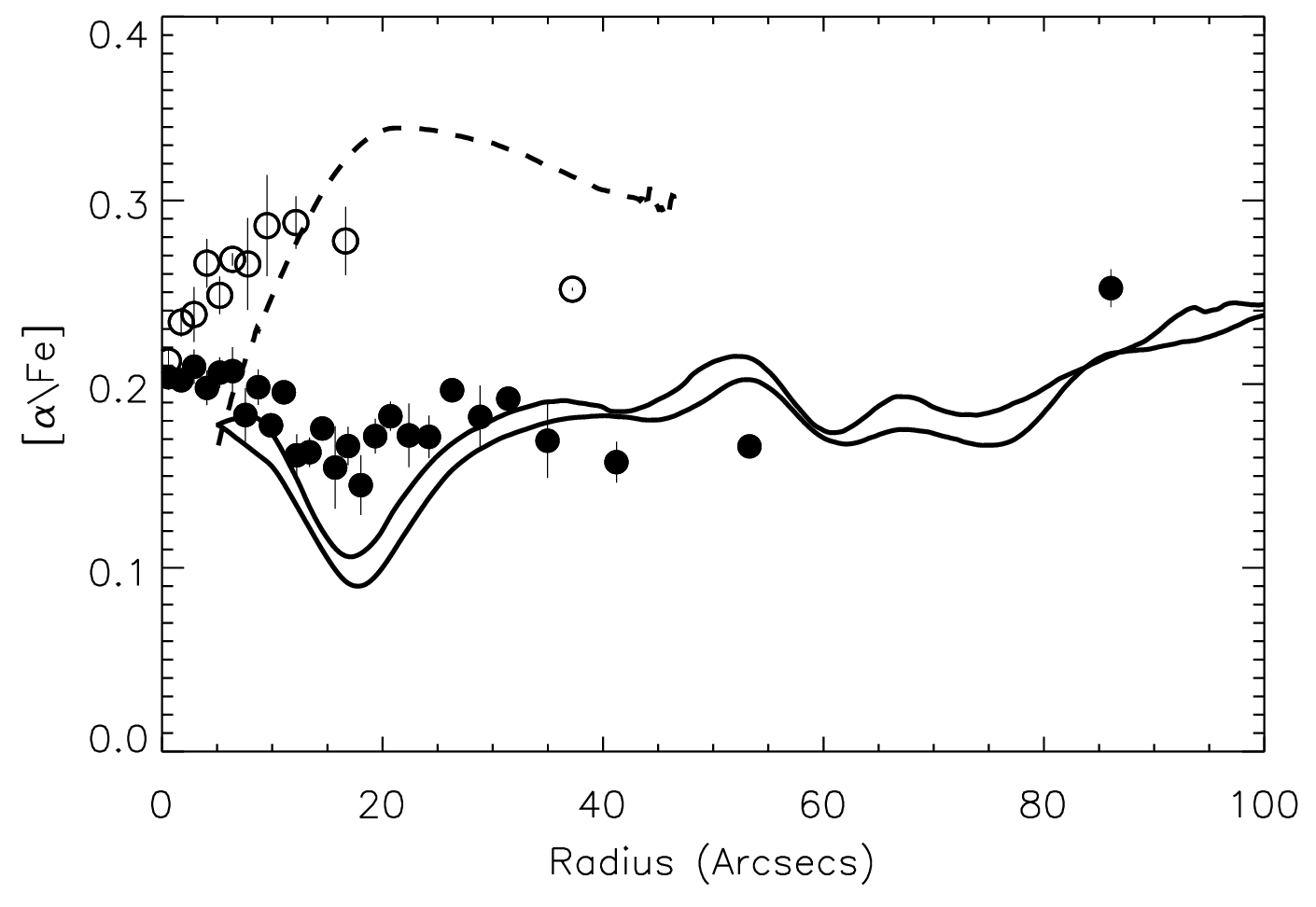} 
   \caption{Model predictions for variation of [$\alpha$/Fe] for major and minor axes.
   Solid line shows the model prediction for the major axis, dashed line 
   displays the model prediction for the minor axis, both made assuming
    [$\alpha$/Fe] = 0.0 for the disc and 0.3 for the spheroid. The circles show radially binned
   [$\alpha$/Fe] measurements shown previously in Fig. 5 for the respective axes. Slight 
   differences between both sides of the major axis are visible in the model predictions, 
   which have been folded about the centre.}
   \label{fig:9}
\end{figure}

\section[]{Discussion}

        In general all of our kinematical measurements are in good agreement with
previous studies, the most interesting finding being that the disc of the galaxy
is particularly cold ($\upsilon$$_{\it{disc}}$ / $\sigma$$_{\it{disc}}$ $>$ 2) and hence
rotationally supported. An interesting extension to this work would be to improve the statistics of the kinematics beyond 100 arcsecs, which would probe the region where
substantial flaring of the disc has been observed.
The fact that the rotation velocity on the major axis remains high even at large radii
where the integrated light is dominated by the underlying spheroid implies that
in NGC 3115 the halo is also rotating significantly (v / $\sigma$ $\sim$ 1.5)
in the same sense as the disc.

        We have confirmed that the GC population shows clear evidence for rotation, 
in the same sense as that of the disc. Statistics on the GC kinematics at 
present are not sufficient to associate particular cluster populations to any 
specific structural feature.

        Our investigation of the Lick absorption line indices has shown that 
the stellar disc component is considerably different from the spheroidal 
component in almost all measured parameters. Most informative in terms 
of constraints on formation theories is the implication that the 
disc of the NGC 3115 is several Gyr younger, and significantly enriched 
in metals, compared to the spheroid of the galaxy. This has been hinted 
at previously by studies of the optical colours of NGC 3115, for example
 \citet{Silva89}, who noted that in B - $\it{i}$ the disc was 0.5 mag bluer than
the spheroid. This colour difference however could be attributed either to a younger 
stellar population in the disc or to the disc having a lower metal
abundance compared to the spheroid. We have convincingly demonstrated that
a lower metal abundance cannot explain this colour difference since the 
major axis displays a $\it{higher}$ $\it{metal}$ $\it{abundance}$ than that of the minor axis.
The difference in colour can therefore clearly be attributed to an age difference
of 5 - 7 Gyr for the two components and is also consistent with the observation
that [$\alpha$/Fe] ratios are lower on the inner region of the major axis.

        The red GC sub-population is most consistent with the larger radii minor axis data 
in its measured parameters. This suggests that both the spheroid and the red GC 
population may have formed from the same material at around the same time. An 
interesting extension would be to probe to larger radii to see if the observed trends 
in metallicity and [$\alpha$/Fe] shown in Fig. 5 continue, and eventually
lead to the minor axis spheroid data tracing stellar populations similar to those that make
up the lower metallicity blue GC population. This blue population of GC's could then possibly be
attributed to an initial burst of star formation during the halo formation of the galaxy.
 One unanswered question is whether or not there exists a GC population associated with the substantial younger disc component of this galaxy. At present the number of GC's for which
 spectroscopy exists is too small to exclude this possibility. Though at present none of the
 GC's examined have line-strength indices comparable to those measured for the disc component.
 If no such population exists, this in 
 itself would prove interesting as it would imply that GC's are not necessarily formed 
 whenever a sizeable amount of star formation occurs.

\section[]{Conclusions}

        We have presented new kinematical data and absorption line strength 
measurements for the major and minor axes of the S0 galaxy NGC 3115, and 
have compared these measurements to similar data for the GC system of
NGC 3115.
        Our main conclusions are:

\begin{itemize}

\item   NGC 3115 has a significant stellar disc component, which is
         both kinematically and chemically distinct from the surrounding
        spheroidal component.\\

\item   The spheroidal component of NGC 3115 is consistent
        with having a uniformly old $\sim$ 10 - 12 Gyr age and [$\alpha$/Fe] of 
        0.2 - 0.3. At large radii the minor axis (which should trace
        the spheroidal component exclusively beyond the central few arcsec) 
        is consistent in age, [$\alpha$/Fe] and metallicity with the red GC 
        sub-population, hinting at a common origin for the two.\\

\item   The major axis data displays clear evidence for contamination by a
        younger (5 - 8 Gyr old) more chemically enriched stellar disc. The 
        observation that the disc of NGC 3115 is bluer than 
        the spheroid is primarily an age difference not a metallicity difference effect.\\

\item Previously observed discrepancies in age determination between
        the H$\beta$ and higher order Balmer lines for the GC sample can 
        largely be explained by changes in the higher 
        order Balmer lines due to varying [$\alpha$/Fe].\\

\item   The GC system displays clear evidence for prograde rotation in the same
        sense as the disc and spheroidal components.

\end{itemize}

\section*{Acknowledgments}
The authors would like to thank Mark Swinbank and Jim Geach for
their useful input and discussions. We also thank the anonymous referee
for several suggestions which improved the presentation of this paper.
MAN acknowledges financial support from PPARC.

Based on observations obtained at the Gemini Observatory, which is operated by the
Association of Universities for Research in Astronomy, Inc., under a cooperative agreement
with the NSF on behalf of the Gemini partnership: the National Science Foundation (United
States), the Particle Physics and Astronomy Research Council (United Kingdom), the
National Research Council (Canada), CONICYT (Chile), the Australian Research Council
(Australia), CNPq (Brazil) and CONICET (Argentina)

\bibliographystyle{mn2e}
\bibliography{references}

\begin{thebibliography}{}

\bibitem[\protect\citeauthoryear{{Bender}, {Saglia} \& {Gerhard}}{{Bender}
  et~al.}{1994}]{Bender94}
{Bender} R.,  {Saglia} R.~P.,    {Gerhard} O.~E.,  1994, \mnras, 269, 785

\bibitem[\protect\citeauthoryear{{Capaccioli}, {Cappellaro}, {Held} \&
  {Vietri}}{{Capaccioli} et~al.}{1993}]{Capaccioli93}
{Capaccioli} M.,  {Cappellaro} E.,  {Held} E.~V.,    {Vietri} M.,  1993, \aap,
  274, 69

\bibitem[\protect\citeauthoryear{{Capaccioli}, {Vietri} \& {Held}}{{Capaccioli}
  et~al.}{1988}]{Capaccioli88}
{Capaccioli} M.,  {Vietri} M.,    {Held} E.~V.,  1988, \mnras, 234, 335

\bibitem[\protect\citeauthoryear{{Cappellari} \& {Emsellem}}{{Cappellari} \&
  {Emsellem}}{2004}]{pPXF}
{Cappellari} M.,  {Emsellem} E.,  2004, \pasp, 116, 138

\bibitem[\protect\citeauthoryear{{Elson}}{{Elson}}{1997}]{Elson97_n3115}
{Elson} R.~A.~W.,  1997, \mnras, 286, 771

\bibitem[\protect\citeauthoryear{{Emsellem}, {Dejonghe} \& {Bacon}}{{Emsellem}
  et~al.}{1999}]{Emsellem99}
{Emsellem} E.,  {Dejonghe} H.,    {Bacon} R.,  1999, \mnras, 303, 495

\bibitem[\protect\citeauthoryear{{Fisher}}{{Fisher}}{1997}]{Fisher97}
{Fisher} D.,  1997, \aj, 113, 950

\bibitem[\protect\citeauthoryear{{Fisher}, {Franx} \& {Illingworth}}{{Fisher}
  et~al.}{1996}]{Fisher96}
{Fisher} D.,  {Franx} M.,    {Illingworth} G.,  1996, \apj, 459, 110

\bibitem[\protect\citeauthoryear{{Gonz{\' a}lez}}{{Gonz{\'
  a}lez}}{1993}]{Gonzalez93}
{Gonz{\' a}lez} J.~J.,  1993, Ph.D.~Thesis,~University of California,~Santa
  Cruz

\bibitem[\protect\citeauthoryear{{Hook}, {J{\o}rgensen}, {Allington-Smith},
  {Davies}, {Metcalfe}, {Murowinski} \& {Crampton}}{{Hook}
  et~al.}{2004}]{Hook04}
{Hook} I.~M.,  {J{\o}rgensen} I.,  {Allington-Smith} J.~R.,  {Davies} R.~L.,
  {Metcalfe} N.,  {Murowinski} R.~G.,    {Crampton} D.,  2004, \pasp, 116, 425

\bibitem[\protect\citeauthoryear{{Illingworth} \& {Schechter}}{{Illingworth} \&
  {Schechter}}{1982}]{Illingworth82}
{Illingworth} G.,  {Schechter} P.~L.,  1982, \apj, 256, 481

\bibitem[\protect\citeauthoryear{{Jones}}{{Jones}}{1997}]{Jones97}
{Jones} L.~A.,  1997, Ph.D thesis, Univ. North Carolina, Chapel Hill

\bibitem[\protect\citeauthoryear{{Kavelaars}}{{Kavelaars}}{1998}]{Kavelaars}
{Kavelaars} J.~J.,  1998, \pasp, 110, 758

\bibitem[\protect\citeauthoryear{{Kormendy} \& {Richstone}}{{Kormendy} \&
  {Richstone}}{1992}]{Kormendy92}
{Kormendy} J.,  {Richstone} D.,  1992, \apj, 393, 559

\bibitem[\protect\citeauthoryear{{Kundu} \& {Whitmore}}{{Kundu} \&
  {Whitmore}}{1998}]{Kundu98}
{Kundu} A.,  {Whitmore} B.~C.,  1998, \aj, 116, 2841

\bibitem[\protect\citeauthoryear{{Kuntschner}}{{Kuntschner}}{2000}]{2000MNRAS.%
315..184K}
{Kuntschner} H.,  2000, \mnras, 315, 184

\bibitem[\protect\citeauthoryear{{Kuntschner}}{{Kuntschner}}{2004}]{Kuntschner%
04}
{Kuntschner} H.,  2004, \aap, 426, 737

\bibitem[\protect\citeauthoryear{{Kuntschner}, {Ziegler}, {Sharples}, {Worthey}
  \& {Fricke}}{{Kuntschner} et~al.}{2002}]{Kuntschner02}
{Kuntschner} H.,  {Ziegler} B.~L.,  {Sharples} R.~M.,  {Worthey} G.,
  {Fricke} K.~J.,  2002, \aap, 395, 761

\bibitem[\protect\citeauthoryear{{Michard} \& {Marchal}}{{Michard} \&
  {Marchal}}{1994}]{Michard_n3115}
{Michard} R.,  {Marchal} J.,  1994, \aaps, 105, 481

\bibitem[\protect\citeauthoryear{{Peng}, {Ho}, {Impey} \& {Rix}}{{Peng}
  et~al.}{2002}]{Peng02_Galfit}
{Peng} C.~Y.,  {Ho} L.~C.,  {Impey} C.~D.,    {Rix} H.-W.,  2002, \aj, 124, 266

\bibitem[\protect\citeauthoryear{{Proctor}, {Forbes} \& {Beasley}}{{Proctor}
  et~al.}{2004}]{2004MNRAS.355.1327P}
{Proctor} R.~N.,  {Forbes} D.~A.,    {Beasley} M.~A.,  2004, \mnras, 355, 1327

\bibitem[\protect\citeauthoryear{{Puzia}, {Kissler-Patig}, {Thomas},
  {Maraston}, {Saglia}, {Bender}, {Goudfrooij} \& {Hempel}}{{Puzia}
  et~al.}{2005}]{Puzia05}
{Puzia} T.~H.,  {Kissler-Patig} M.,  {Thomas} D.,  {Maraston} C.,  {Saglia}
  R.~P.,  {Bender} R.,  {Goudfrooij} P.,    {Hempel} M.,  2005, \aap, 439, 997

\bibitem[\protect\citeauthoryear{{Puzia}, {Kissler-Patig}, {Thomas},
  {Maraston}, {Saglia}, {Bender}, {Richtler}, {Goudfrooij} \& {Hempel}}{{Puzia}
  et~al.}{2004}]{Puzia04}
{Puzia} T.~H.,  {Kissler-Patig} M.,  {Thomas} D.,  {Maraston} C.,  {Saglia}
  R.~P.,  {Bender} R.,  {Richtler} T.,  {Goudfrooij} P.,    {Hempel} M.,  2004,
  \aap, 415, 123

\bibitem[\protect\citeauthoryear{{Puzia}, {Zepf}, {Kissler-Patig}, {Hilker},
  {Minniti} \& {Goudfrooij}}{{Puzia} et~al.}{2002}]{Puzia02}
{Puzia} T.~H.,  {Zepf} S.~E.,  {Kissler-Patig} M.,  {Hilker} M.,  {Minniti} D.,
     {Goudfrooij} P.,  2002, \aap, 391, 453

\bibitem[\protect\citeauthoryear{{Schlegel}, {Finkbeiner} \&
  {Davis}}{{Schlegel} et~al.}{1998}]{Schlegel1998}
{Schlegel} D.~J.,  {Finkbeiner} D.~P.,    {Davis} M.,  1998, \apj, 500, 525

\bibitem[\protect\citeauthoryear{{Silva}, {Boroson}, {Thompson} \&
  {Jedrzejewski}}{{Silva} et~al.}{1989}]{Silva89}
{Silva} D.~R.,  {Boroson} T.~A.,  {Thompson} I.~B.,    {Jedrzejewski} R.~I.,
  1989, \aj, 98, 131

\bibitem[\protect\citeauthoryear{{Thomas}, {Maraston} \& {Bender}}{{Thomas}
  et~al.}{2003}]{Thomas03}
{Thomas} D.,  {Maraston} C.,    {Bender} R.,  2003, \mnras, 339, 897

\bibitem[\protect\citeauthoryear{{Thomas}, {Maraston} \& {Korn}}{{Thomas}
  et~al.}{2004}]{Thomas04}
{Thomas} D.,  {Maraston} C.,    {Korn} A.,  2004, \mnras, 351, L19

\bibitem[\protect\citeauthoryear{{Trager}, {Worthey}, {Faber}, {Burstein} \&
  {Gonzalez}}{{Trager} et~al.}{1998}]{Trager98}
{Trager} S.~C.,  {Worthey} G.,  {Faber} S.~M.,  {Burstein} D.,    {Gonzalez}
  J.~J.,  1998, \apjs, 116, 1

\bibitem[\protect\citeauthoryear{{Vazdekis}}{{Vazdekis}}{1999}]{Vazdekis99}
{Vazdekis} A.,  1999, \apj, 513, 224

\bibitem[\protect\citeauthoryear{{Worthey} \& {Ottaviani}}{{Worthey} \&
  {Ottaviani}}{1997}]{Worthey97}
{Worthey} G.,  {Ottaviani} D.~L.,  1997, \apjs, 111, 377

\end{thebibliography}
\label{lastpage}

\end{document}